\newcommand{\CLASSINPUTtoptextmargin}{0.75in}
\newcommand{\CLASSINPUTbottomtextmargin}{0.9in}
\newcommand{\CLASSINPUToutersidemargin}{1.65cm}

\documentclass[10pt,twocolumn,conference]{IEEEtran}

\usepackage{amsmath, mathrsfs, mathtools}
\usepackage{amsfonts}
\usepackage{amssymb}
\usepackage{amsthm}
\usepackage{dsfont}
\usepackage{graphicx}
\usepackage{color}
\usepackage{graphicx}
\usepackage{subfig}
\usepackage{caption}
\usepackage[backend=bibtex,giveninits=true,sorting=none]{biblatex}
\usepackage{balance}
\renewcommand*{\bibfont}{\footnotesize}

\let\labelindent\relax
\usepackage{enumitem}

\newcommand{\subparagraph}{}
\usepackage[compact]{titlesec}
\titlespacing*{\section}{15pt}{1.2\baselineskip}{0.9\baselineskip}

\theoremstyle{remark}

\long\def\comment#1{}

\newcommand\figref{Figure~\ref}

\newcommand{\ben}{\begin{enumerate}}
\newcommand{\een}{\end{enumerate}}

\newcommand{\beq}{\begin{equation}}
\newcommand{\eeq}{\end{equation}}

\newcommand{\bi}{\begin{itemize}}
\newcommand{\ei}{\end{itemize}}

\newcommand{\CC}{\mathbb{C}}

\newcommand{\RR}{\mathbb{R}}

\newcommand{\EE}{\mathbb{E}}

\newcommand{\av}{{\bf a}}

\newcommand{\cv}{{\bf c}}

\newcommand{\ev}{{\bf e}}

\newcommand{\hv}{{\bf h}}

\newcommand{\sv}{{\bf s}}

\newcommand{\xv}{{\bf x}}
\newcommand{\yv}{{\bf y}}
\newcommand{\zv}{{\bf z}}

\newcommand{\Am}{{\bf A}}
\newcommand{\Bm}{{\bf B}}
\newcommand{\Cm}{{\bf C}}

\newcommand{\Hm}{{\bf H}}
\newcommand{\Id}{{\bf I}}

\newcommand{\Sm}{{\bf S}}

\newcommand{\Wm}{{\bf W}}

\newcommand{\Xm}{{\bf X}}
\newcommand{\Ym}{{\bf Y}}
\newcommand{\Zm}{{\bf Z}}

\newcommand{\Cc}{{\cal C}}

\newcommand{\Kc}{{\cal K}}
\newcommand{\Lc}{{\cal L}}

\newcommand{\Gammam}{\boldsymbol{\Gamma}}

\newcommand{\SINR}{{\sf SINR}}
\newcommand{\SNR}{{\sf SNR}}

\newcommand{\herm}{{\sf H}}

\title{Pilot-Based Unsourced Random Access with a Massive MIMO Receiver in the Quasi-Static Fading Regime}
\author{Alexander Fengler, Peter Jung, Giuseppe Caire
\thanks{The authors are with the Communications and Information Theory Group,
Technische Universit\"{a}t Berlin (\{fengler, peter.jung,
caire\}@tu-berlin.de).}
}

\begin{document}

\maketitle

\begin{abstract}
    In this work we treat the unsourced random access problem on a Rayleigh block-fading AWGN channel with
    multiple receive antennas. Specifically, we consider the slowly fading scenario where the coherence
    block-length is large compared to the number of active users and the message can be transmitted in one coherence block.
    Unsourced random access refers to 
    a form of grant-free random access where users are considered to be a-priori indistinguishable and 
    the receiver recovers a list of transmitted messages up to permutation.
    In this work we show that, when the coherence block length is large enough, a conventional approach
    based on the transmission of non-orthogonal pilot sequences with subsequent channel estimation and Maximum-Ratio-Combining (MRC)
    provides a simple energy-efficient solution whose performance can be well approximated in closed form.
    For the finite block-length simulations we use a randomly sub-sampled DFT matrix as pilot matrix,
    a low-complexity approximate message passing algorithm for activity detection and
    a state-of-the-art polar code with a successive-cancellation-list decoder as single-user error correction code.
    These simulations prove the scalability of the presented approach and the quality of the analysis.
\end{abstract}

\begin{keywords}
    Internet of Things (IoT), Unsourced Random Access, Massive Multi-User MIMO, Approximate Message Passing (AMP).
\end{keywords}

\section{Introduction}
Conventional random-access protocols in current mobile communication standards
establish an uplink connection between user and base station (BS) by first
running a multi-stage handshake protocol \cite{%
,Dah2018}.
During this initial access phase active users are identified and subsequently
a scheduler assigns orthogonal transmission resources to the active users.
One of the paradigms of modern machine-type communications
consists of a very large
number of devices (here referred to as ``users'') with sporadic data. Typical examples thereof are
Internet-of-Things (IoT) applications, wireless sensors deployed to monitor smart infrastructure,
and wearable biomedical devices \cite{Has2013}. In such scenarios, a BS should be
able to collect data from a large number of devices. However, due to the sporadic nature of the
data generation and communication, an initial access procedure is overly wasteful.

An alternative way of communication is known as grant-free random-access, where users transmit their
data without awaiting the grant of transmission resources by the BS. A commonly discussed 
grant-free strategy in a massive multi-user MIMO setting is to assign 
fixed orthogonal or non-orthogonal pilot sequences to users
\cite{Lar2014,Mar2016,Liu2018a,Liu2018b,Che2018,Liu2018c,Sen2018b}. Active users then transmit their pilot sequence
directly followed by their data sequence.
The BS identifies the active users in the first step and estimates their channel vectors. Subsequently
the channel estimates are used to detect the data sequences using either maximum-ratio-combining (MRC)
or zero-forcing \cite{Mar2016}.

However, as the number of users in a system grows large and the access frequency becomes small, it gets
increasingly inefficient to assign fixed pilot sequences to all the users.
In contrast, {\em unsourced random access} (U-RA) is a novel grant-free paradigm proposed in \cite{Pol2017}
and motivated by an IoT scenario where millions of cheap devices have their codebook hardwired at
the moment of production, and are then disseminated into the environment.  In this case,  all users
make use of the very same codebook and the BS decodes the list of transmitted messages
irrespectively of the identity of  of the active users.
The U-RA approach can simplify the random-access protocol design because it does not required an initial
access phase and, in contrast to existing grant-free approaches, it allows for a system that is completely independent of the inactive users, which makes
it well suited to the IoT scenario with a huge number of devices with very sporadic activity. 

Practical coding schemes for U-RA have been mainly studied on the real AWGN channel,
e.g. \cite{Pol2017,Fen2019c,Fen2020c,Vem2017,Kow2019c,Pra2020b,Ust2019},
and on the Rayleigh quasi-static fading AWGN channel \cite{Kow2020,And2020a}.  
The U-RA problem on a Rayleigh block-fading AWGN channel in a massive MU-MIMO setting was formulated
in \cite{Fen2021a} and it was shown that a covariance-based activity detection (AD) algorithm combined with a tree code \cite{Vem2017}
can achieve sum-spectral-efficiencies that grow proportional to the coherence block-length $n$,
even if the number of active users is significantly larger then $n$,
specifically up to $K_a = \mathcal{O}(n^2)$. 

In typical wireless systems the coherence block-length $n$ may range from
a couple of hundred to a couple
of thousand, depending mainly on the speed of the transmitters. At a carrier frequency of
$2$ GHz the coherence times, according to the model $T_c \approx 1/(4 D_s)$ where $D_s$ is the maximal Doppler  
spread \cite{Tse2005}, may range from 
$45$ ms at $3$ km/h to $1$ ms at $120$ km/h. The coherence bandwidth depends on the maximal delay spread and,
in an outdoor environment,
typically ranges from
$100$ to $500$ kHz depending on the propagation conditions.
Therefore, the number of complex symbols in an OFDM coherence block
may range from $n=100$ to $n=20000$, depending mainly on the assumed speed and the geometry of the environment.
Unfortunately, the covariance-based AD algorithm in \cite{Fen2021a} has a run-time complexity
that scales with $n^2$, which makes it unfeasible to use at $n \gtrsim 300$. So while the covariance based approach
of \cite{Fen2021a} is well suited to a fast-fading scenario with $n\lesssim 300$, it becomes unfeasible for
large coherence block-lengths in the order of thousands. On the other hand,
algorithms based on hybrid-GAMP \cite{Shy2020a} and tensor-based modulation \cite{Dec2020c}
have shown an excellent performance
at large coherence block-length at which the covariance-based approach is no longer feasible. 

In this work we present a conceptually simple algorithm that can be used when $n>K_a$. It is based
on pilot transmission, AD, channel estimation, MRC and single-user decoding, very similar
to the state-of-the art approach for massive MIMO grant-free random access \cite{Liu2018a,Liu2018b,Sen2018b}.
In contrast to the scheme with fixed pilots allocated to all users, we use a pool of non-orthogonal pilots
from which active users pick one pseudo-randomly based on the first bits of their message.
We show that a collision of users, i.e. two users picking the same pilot sequence,
can be resolved by using a polar single-user code with a successive-cancellation-list (SCL) decoder.  
Note, that multiuser MRC detection and single-user polar decoding are completely separated, i.e.
succesive cancellation refers purely to the single-user decoding technique \cite{Bal2015}. 
Finite-length simulations shows that the performance of the coding scheme can be well predicted by analytical
calculations. Despite its simplicity the suggested scheme has an energy efficiency that is comparable to
existing approaches. 
Note, that the problem treated here is mathematically
almost equivalent to grant-free random-access with fixed pilots allocated to each user. Differences arise only
in the possibility of collisions and the associated use of a list decodable single-user code. 
The error probability of AD and MRC in the asymptotic limit $K_a,K_\text{tot},n \to \infty$ with fixed ratios $K_a/K_\text{tot}$ and
$K_\text{tot}/n$ has been analysed in \cite{Liu2018b}. In this work we focus on the finite-blocklength regime
and the combination of MRC with a single-user polar code.
\section{Channel model}
\label{sec:model}
We consider a quasi-static Rayleigh fading channel with 
a block of $n$ signal dimensions over which the user channel vectors are constant.
In contrast to the block-fading channel treated in \cite{Fen2021a}, were a message is encoded over multiple independent
fading blocks, here we assume that a message can be transmitted
in a single coherence block. 
Following the problem formulation in \cite{Pol2017}, each user is given the same codebook
$\Cc = \{ \cv(m) : m \in [2^{nR}]\}$, formed by 
$2^{nR}$ codewords $\cv(m) \in \CC^n$. 
The codewords are normalized such that $\|\cv(m)\|^2_2 \leq n$.
An unknown number $K_a$ out of $K_{\rm tot}$ total users transmit their
message over the coherence block.
Let $\mathcal{K}_a$ denote the set of active users, $i_k$ denote the index of the message chosen by user $k$,
$h_{k,m}\sim\mathcal{CN}(0,1)$ iid be the Rayleigh channel coefficient between user $k$ and receive antenna $m$
and let $g_k\in\RR_+$ denote the large-scale fading coefficient (LSFC) of user $k$, which captures the path-loss
and shadowing components of the fading. Furthermore, let $\gamma_k = g_k$ for $k\in \mathcal{K}_a$ and
zero otherwise. The received signal at the $m$-th receive antenna takes the form
\beq
\yv_m =\sum_{k=1}^{K_\text{tot}}\sqrt{P\gamma_k}h_{k,m} \cv_{i_k} + \zv_m = \sum_{k\in\mathcal{K}_a} \sqrt{P g_k} h_{k,m} \cv_{i_k} + \zv_m 
\eeq 
where $z_{m,i}\sim\mathcal{CN}(0,N_0)$ iid.  
The BS must then produce a list $\Lc$ of the transmitted messages $\{m_k : k\in \Kc_a\}$ (i.e., 
the messages of the active users). Let $n_\text{md} = \sum_{k \in \mathcal{K}_a} \mathds{1}_{\{m_k \notin \mathcal{L}\}}$ denote the number
of transmitted messages missing in the output list.
The system performance is expressed in terms of the {\em Per-User Probability of Misdetection}, defined as
$p_{md} = \EE\left\{ n_\text{md}  \right\}/K_a$.
In applications a slight overhead in the list size may be tolerable if it reduces the misdetections.
Since the number of active users is not necessarily known, it is practical to let the decoder decide
on a list size, which therefore becomes a random variable. 
Let $n_\text{fa} = |\mathcal{L}\setminus \{ m_k : k \in \Kc_a \}|$
denote the number of messages in the output list that were not transmitted by any user, also called
\emph{False Alarms}.
$n_\text{fa}$ is related to the list size by
\beq
|\mathcal{L}| = n_\text{fa} + K_a - n_\text{md}
\label{eq:fa_md_relation}
\eeq 
To get an empirical performance measure we define  
the {\em Probability of False-Alarm} as the average fraction of false alarms, i.e., 
\beq
p_{fa} = \EE\left\{\frac{n_\text{fa}}{|\mathcal{L}|}\right\}.
\label{eq:ura_pfa}
\eeq
Operationally, $p_\text{fa}$ is the probability that a randomly chosen message from the output list
is a false alarm.
In the special case where $K_a$ is known at the receiver and $|\mathcal{L}| = K_a$ is fixed it follows from \eqref{eq:fa_md_relation}
that $p_\text{fa} = p_\text{md}$.

Notice that in this problem formulation the number of total users $K_{\rm tot}$ is completely irrelevant,
as long as it is much larger than the range of possible active user set sizes $K_a$
(e.g., we may consider $K_{\rm tot} = \infty$). 
Furthermore, as
customary in coded systems, we express energy efficiency in terms of the
standard quantity $E_b/N_0 :=  \frac{P}{R N_0}$. 

In line with the classical massive MIMO setting \cite{Mar2016},
we assume an independent Rayleigh fading model for the channel coefficients $h_{k,m}$,
such that the channel vectors
for different users are independent from each other and are spatially white (i.e.,
uncorrelated along the antennas), that is,  $\hv_k = (h_{k,1},...,h_{h,m})^\top \sim \mathcal{CN}(0, \Id_M)$.

\section{Pilot-based massive MIMO U-RA}
Let the coherence block be divided into two periods of lengths $n_p$ and $n_d$.
In the first period each users chooses one of $N = 2^J$ (non-orthogonal) pilot sequences based on the first
$J$ bits of its message. The received signal in this phase is given by
\begin{align}\label{pilot_sig}
    \Ym_p = \Am \Gammam^{\frac{1}{2}} \Hm + \Zm_p \in \CC^{n_p \times M},
\end{align}
where $\Am \in \CC^{n_p\times N}$ is the matrix of pilot sequences with columns normalized as $\|\av_i\|_2^2 = n_p$,
$\Hm = (\hv_1,...,\hv_{N})^\top \in \CC^{N\times M}$ is the matrix with the channel
vectors as rows
and $\Gammam$ is the matrix with $P_\text{pilot}\gamma_k$ on the diagonal.
Note, that channel vectors are only defined for those indices which have been chosen by the active users.
Formally, we define the remaining channel vectors as zero.
The BS uses an AD algorithm as in \cite{Fen2021a} to estimate the indices of
the used pilots and the corresponding LSFCs.
Let $\hat{\Gammam}$ denote the matrix with the estimates $\hat{\gamma}_k$ of $\gamma_k$ on the diagonal and let 
$\hat{\mathcal{I}}$ be the estimate of the set of active users. 
$\hat{\mathcal{I}}$ can be obtained by either thresholding the estimated received powers, i.e.
\beq
\hat{\mathcal{I}} = \{k: P_\text{pilot}\hat{\gamma}_k > \theta\}
\eeq 
where $\theta$ is some suitable threshold, or, if the number of active users is available,
by picking the indices corresponding to the $K_a$ largest estimated received powers.
Also for an index set $\mathcal{I}$ and for any matrix $\Bm$ let $\Bm_\mathcal{I}$
denote the matrix that contains only the columns of $\Bm$ with indices in $\mathcal{I}$. 
Then a linear MMSE estimate of the channel matrix is computed as
\beq
\hat{\Hm} = \hat{\Gammam}_{\hat{\mathcal{I}}}^{1/2}\Am_{\hat{\mathcal{I}}}^\herm \left(\Am_{\hat{\mathcal{I}}}\hat{\Gammam}_{\hat{\mathcal{I}}}\Am_{\hat{\mathcal{I}}}^\herm + N_0\Id_{n_p}\right)^{-1}\Ym_p \in \CC^{\hat{K}_a \times M}
\label{eq:lmmse}
\eeq 
where $\Am_{\hat{\mathcal{I}}}$ denotes a sub-matrix of the pilot matrix $\Am$ which contains only the columns
which have been estimated as active and $\hat{\Gammam}_{\hat{\mathcal{I}}}$ contains the
estimated LSFCs $\hat{\gamma}_k$ of the active users on the diagonal. 
In the second period each users encodes its remaining $B-J$-bit message with a binary $(B-J,2n_d)$
block code and modulates
the $2n_d$ coded bits via QPSK on a sequence of $n_d$ complex symbols $\sv_k$.
These are transmitted over the $n_d$ channel uses
in the second phase. The matrix of received signals in the second phase is
\beq 
\Ym_d = \sum_{k\in\mathcal{K}_a} \sqrt{P_\text{data}g_k}\sv_k\hv_k + \Zm_d \in \CC^{n_d \times M}.
\eeq 
The BS uses the channel estimate $\hat{\Hm}$ from the first phase to perform multiuser detection via
MRC,
i.e. it computes 
\beq
\hat{\Sm} = \hat{\Gammam}_{\hat{\mathcal{I}}}^{-1/2}\hat{\Hm}\Ym_d^\herm \in \CC^{\hat{K}_a \times n_d}
\eeq 
The rows of $\hat{\Sm}$ correspond to estimates of the transmitted sequences $\sv_k$.
Note, that it is also possible to use zero-forcing \cite{Mar2016}
instead of MRC but this would require that $M>K_a$.
The rows of $\hat{\Sm}$ are individually demodulated,
the bit-wise log-likelihood ratios are computed and fed into a soft-input single-user decoder. 
If the decoder finds a valid codeword, the index of the corresponding pilot is converted back to a
$J$ bit sequence and prepended to the codeword. Then the combination of the two is added to the output list.
The use of a polar code with CRC-bits and a SCL decoder has the additional benefit
that we can include all the valid codewords in the output list of the SCL decoder in the U-RA output list.
This allows to recover the messages of colliding users which have chosen the same pilot in the first phase.
The ability of polar codes to resolve sums of codewords has been observed and used for U-RA on the AWGN
in combination with spreading sequences \cite{Pra2020b} and a slotted Aloha approach \cite{Mar2019,And2020a}.
\subsection{Activity Detection}
\label{sec:mmv_amp}

For the AD in the pilot phase we use the MMV-AMP algorithm, which was
introduced in \cite{Kim2011}, and used for 
AD in a Bayesian setting, 
where the LSFCs are either known, or its distribution is known in \cite{Liu2018a,Che2018}. 
The algorithm aims to recover the unknown matrix
$\Xm = \Gammam^{1/2}\Hm$
from the linear Gaussian measurements $\Ym_p$. 
Details are omitted due to space constraints and can be found in \cite{Liu2018a,Che2018, Fen2021a}.
\subsection{DFT Pilots and Fast MMV-AMP}
\label{sec:fast_mmv_amp}
To avoid the $\mathcal{O}(Mn_pN)$ complexity of the matrix multiplications
we create the pilot matrix by choosing a random subset of $n_p$ rows of a $N \times N$
DFT matrix. This allows to replace matrix multiplications of the form
$\Am\Xm$ and $\Am^\herm\Zm$ in the MMV-AMP algorithm
by FFT operations.
Let $\mathcal{S} = (s_1,...,s_{n_p}) \subset [N]$ be a randomly chosen subset.
Let $\Wm \in \CC^{N \times N}$ be a DFT matrix defined by
\beq
    W_{i,j} = \omega^{jk}
\eeq 
where $\omega = e^{-2\pi i/N}$. Then we define the pilot matrix
$\Am^\text{DFT} \in \CC^{n_p \times N}$ as the submatrix of $\Wm$ with rows defined as
$\Am^\text{DFT}_{i,:} = \Wm_{s_i,:}$.
The matrix multiplication $\Am^\text{DFT}\xv$ for some arbitrary vector $\xv \in \CC^{N}$
is give by 
\beq 
(\Am^\text{DFT}\xv)_i = (\text{FFT}_{N}(\xv))_{s_i}
\eeq
for $i=1,...,n_p$, where $\text{FFT}:\CC^{N}\to\CC^{N}$ denotes a properly normalized fast-Fourier transform (FFT) operation.
For 
the hermitian transpose matrix multiplication $\Am^{\text{DFT},\herm} \zv$ for some vector $\zv \in \CC^{n_p}$
first define $\tilde{\zv} \in \CC^{N}$ by $\tilde{z}_{s_i} = z_i$ and zero otherwise. Then
\beq
\Am^{\text{DFT},\herm} \zv = \text{FFT}(\tilde{\zv})
\eeq
For the matrix-matrix multiplications $\Am^\text{DFT}\Xm$ this process has to be repeated for each column of $\Xm$
leading to a complexity of $\mathcal{O}(MN\log N)$.
\subsection{Analysis}
\label{sec:mrc_analysis}
In this section we calculate an approximate finite-blocklength lower bound on the error probability and
on the energy efficiency of the MRC approach. 
We assume that the AD and LSFC estimation can be done without errors, i.e. $\hat{\Gammam} = \Gammam$.
This gives a lower bound
on the error probability
and 
in the regime where $n_p>K_a$ we expect it to be tight,
as in this regime the AD error rates and the error of the LSFC
estimation are very low \cite{Liu2018a, Fen2021a}. 
For simplicity we consider $P_\text{pilot} = P_\text{data} =: P$ here.
The covariance of the channel estimation error of the LMMSE estimation in \eqref{eq:lmmse} is given by
\beq
\Cm_e = \Id_{K_a} - \Gammam_\mathcal{I}^{1/2}\Am_\mathcal{I}^\herm \left(\Am_\mathcal{I}\Gammam_\mathcal{I}\Am_\mathcal{I}^\herm + N_0\Id_{n_p}\right)^{-1}\Am_\mathcal{I}\Gammam_\mathcal{I}^{1/2}
\label{eq:error_cov}
\eeq 
and the MSE of the channel estimate of user $k\in\mathcal{K}_a$ is given by
$\sigma^2_k := \EE\{|h_{k,m}-\hat{h}_{k,m}|^2\} = (\Cm_e)_{k,k}$.
A typically tight approximation of the effective $\SINR$ of each user after MRC  
is given by \cite{Mar2016,Cai2018}
\beq
\SINR_k = \frac{M(1 - \sigma_k^2) g_k P}{N_0 + \sigma_k^2 g_k P +  \sum_{j=1, j \neq k }^{K_a}   g_j P} 
\eeq 
In addition to calculating $\sigma^2_k$ via \eqref{eq:error_cov},
a closed form lower bound on $\sigma_k^2$ is given by
\beq
\sigma^2_k \geq \frac{N_0}{N_0+n_p P g_k}    
\label{eq:ortho}
\eeq 
which is obtained by assuming orthogonal pilots. 
We investigate this lower bound because the evaluation of \eqref{eq:ortho} is much simpler
then $\eqref{eq:error_cov}$ since it does not require the inversion of a possibly large matrix.
Furthermore, we can evaluate the impact of the non-orthogonality on the channel estimation
by comparing \eqref{eq:ortho} to the true channel estimation errors obtained from \eqref{eq:error_cov}.

An approximation of the achievable rates of a block-code with block-length $2n_d$ and error probability $p_e$
on a real AWGN channel with power $\SINR$ is given by the normal approximation \cite{Pol2010}
\beq
    R \approx 0.5\log(1+\SINR) - \sqrt{\frac{V}{2n_d}}Q^{-1}(p_e)
    \label{eq:normal}
\eeq 
where 
\beq
V = \frac{\SINR}{2}\frac{\SINR+2}{(\SINR+1)^2}\log^2 e
\eeq
and $Q(\cdot)$ is the Q-function. Using the normal approximation we can find the required $\SINR$ to achieve
a certain error probability at a given block-length and then we can find the required input power
to achieve the target \SINR. 

\subsection{Simulations}
\begin{figure}
   \centering
   \includegraphics[width=0.9\linewidth]{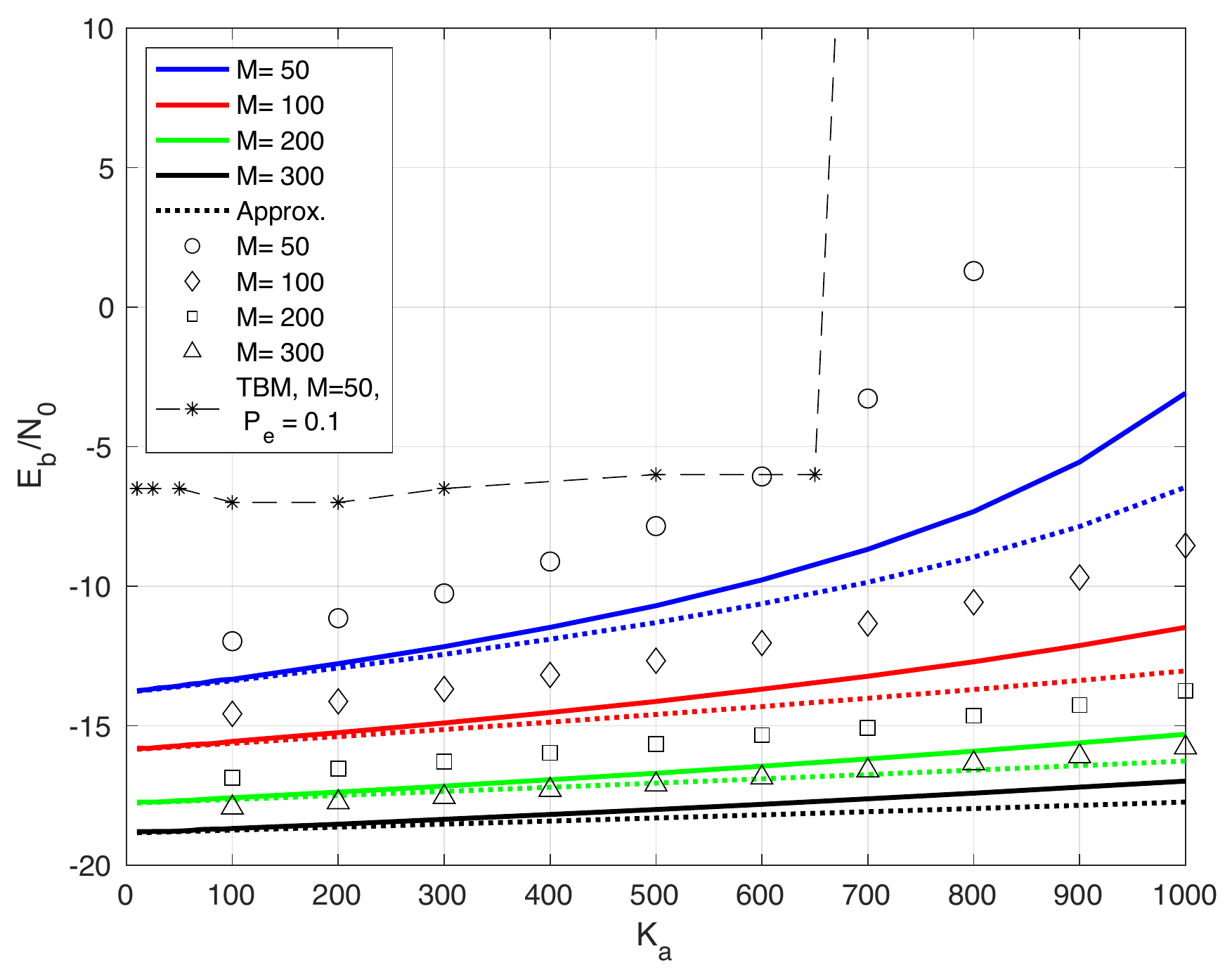}
   \caption[MRC: $E_b/N_0$ over $K_a$]{Required energy-per-bit with the MRC approach to achieve $P_e = p_\text{md} + p_\text{fa} < 0.05$.
  Solid lines represent the achievability estimates from Section \ref{sec:mrc_analysis}. Dotted
  lines represent the theoretical results with the assumption of orthogonal pilots.
  The markers represent simulation results.} 
  \label{fig:ebn0_over_K_mrc}
\end{figure}

For the simulation in \figref{fig:ebn0_over_K_mrc} we choose $n=3200$, $P_e = 0.05$,
$B=100$, $n_p = 1152$, $n_d = 2048$
and $J = 16$.
A randomly sub-sampled DFT matrix is used as pilot matrix and 
in the AD phase we use the MMV-AMP algorithm with an approximate calculation of the derivative
as described in Sections \cite{Fen2021a}.
Furthermore, all LSFCs are considered to be constant $g_k = 1$ and known at the receiver.
For simplicity we assume that $K_a$ is known at the receiver, and after the MMV-AMP iterations
are finished the active columns
are estimated by picking the $K_a$ indices with the largest estimated LSFCs. 
We use a polar code \cite{Ari2009,Bal2015} with a state-of-the-art SCL decoder with
$16$ CRC bits and a list size of $32$. The polar code is constructed by the Bhattacharyya method
\cite{Tal2013a}. The resulting code approaches the normal approximation only up to a $\sim$ 1 dB
gap which explains the discrepancy between the theoretical and the simulation results
in \figref{fig:ebn0_over_K_mrc}.
The curves obtained by using formula \eqref{eq:ortho} for the channel estimation
error provide a rough but usable approximation which gets worse as $K_a$ grows close to $n_p$
(dotted lines). 
For comparison we add the reported values of the tensor-based-modulation (TBM) approach \cite{Dec2020c},
with tensor signature (8,5,5,4,4) and an outer BCH code,
although the values have been obtained with the higher target error rate $P_e = 0.1$.
\subsection{Collisions}
\begin{figure}
   \centering
   \includegraphics[width=0.5\linewidth]{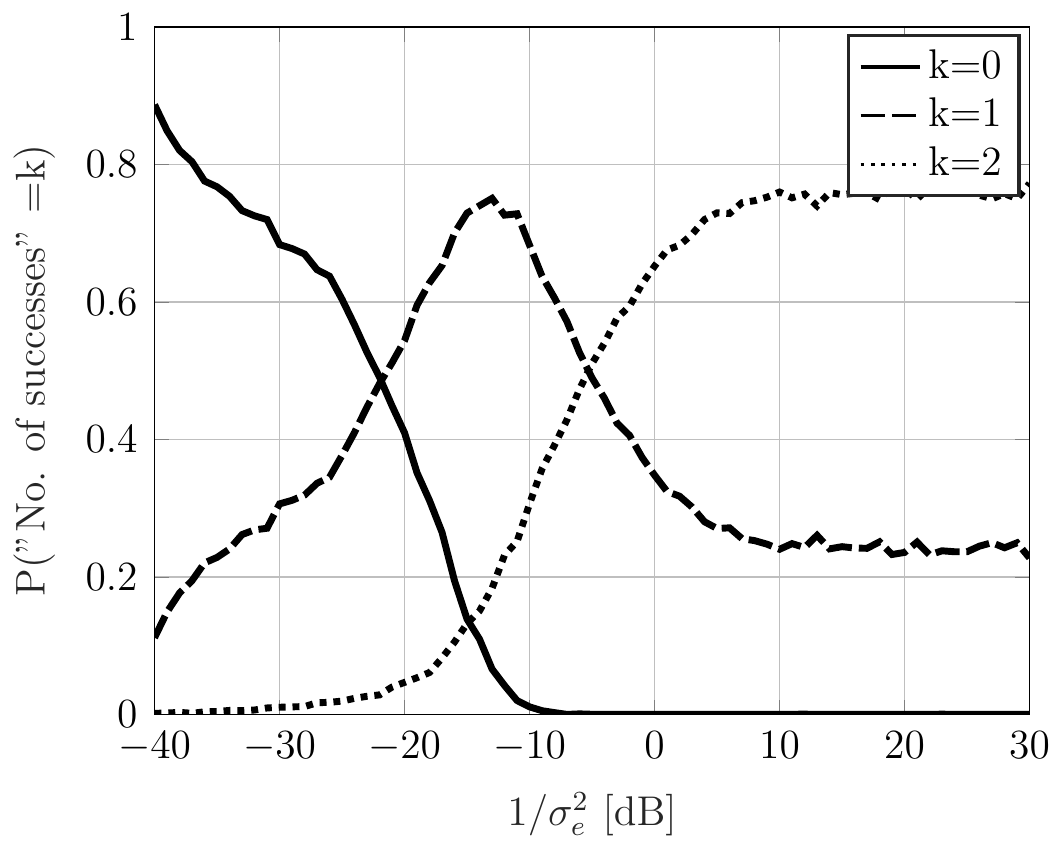}
  \caption{Fractions of correctly recovered codewords in the case of a
      two-user collision with MRC according to the model \eqref{eq:collision_model}.
      Polar code with $B-J=84$ message bits, $n_d=2048$ complex QPSK coded symbols.
}
  \label{fig:collisions_mrc}
\end{figure}

In this paragrapgh we show how pilot collisions affect the decoding performance.
Specificaly, we explain how
the combination of a single-user list decodable code can efficiently
reduce the effect of pilot collisions.
The average number of collisions of $k$ users on one pilot is given by
\beq
\EE\{ C_k\} = \frac{{K_a \choose k}}{N^{k-1}}.   
\label{eq:ev_collisions}
\eeq 
We can safely ignore the collisions of more then two users since their number is much smaller than $1$ for the
considered parameters.
For $K_a = 1000$ and $N=2^{16}$, \eqref{eq:ev_collisions} gives an average number of $7-8$ collisions of order two.
If all of the colliding messages would result in an error, this would lead to a per-user error probability
of $0.016$ on average.
This can be incorporated into the above analysis by subtracting this values from the target
error probability $p_e$ in the normal approximation \eqref{eq:normal}.
Nonetheless, if a list decoder is used as a single-user decoder, it is possible
to recover both of the colliding messages.
This can be explained as follows.
When there is no fading, half of the coded bits are erased on average when two codewords are added.
Since the rate of the polar code
$R = (B-J)/(2n_d)$ is much smaller then $1/2$, these erasure can be recovered, see also \cite{Pra2020b}. 
This situation changes when fading is involved.
In \figref{fig:collisions_mrc} we quantify the effect of pilot collisions 
by taking the uncertainty of the channel estimation and the MRC into account
via the following simplified two user collision model. Let $\sv_1$ and $\sv_2$ denote
the QPSK modulated sequences of two users and $\hv_1,\hv_2 \in \CC^{M}$ their iid Rayleigh
channel vectors. We model the channel estimates as $\hat{\hv} = \hv_1 + \hv_2 + \ev$ where
$\ev \sim \mathcal{CN}(0,\sigma^2_\text{est}\Id_M)$
is the channel estimation error with variance
$\sigma^2_\text{est}$.
The model for the estimated single-user sequence for both users is then
\beq
\hat{\sv}^\top = \frac{1}{M}\hat{\hv}^\herm\left(\sqrt{\SNR}\hv_1\sv^\top_1 + \sqrt{\SNR}\hv_2\sv^\top_2 + \zv\right)
\label{eq:collision_model}
\eeq 
We fix the per-user $\SNR$ to -10 dB and vary the variance of the channel estimation error.
The simulation in \figref{fig:collisions_mrc} shows that for $M=50$
when the channel estimation error $\sigma^2_\text{est}$ falls below
-10 dB both codewords are correctly recovered in around $75\%$ of the cases. 
Nonetheless, the probability that at least one codeword is recovered converges to 1
as the channel estimation error goes to zero.
This effect persist, even when the base per-user $\SNR$ is increased,
because it is a consequence of the cross-correlation terms
$\sqrt{\SNR}\hv_1^\herm\hv_2\sv_2/M$ and $\sqrt{\SNR}\hv_2^\herm\hv_1\sv_1/M$.
However, in the limit $M\to\infty$ these terms vanish and the model \eqref{eq:collision_model}
converges to a 2-user non-fading AWGN unsourced MAC. 

\section{Conclusion}
In this work we presented a coding scheme for the quasi-static Rayleigh fading AWGN
MAC with a massive MIMO receiver that is compliant with the unsourced paradigm.
Simulations show that the presented scheme
performs better than existing approaches for the quasi-static MIMO MAC despite its conceptual
simplicity. We give a closed form analysis that predicts the achievable limits of this scheme when
a single user code is used which can achieve the normal approximation.
The results show that the presented approach can scale to many hundreds of users and achieve
high sum-spectral efficiencies with low power requirement and without any
coordination between users. 
Specifically, with $M=100$ receive antennas over a thousand users
can be served concurrently,
which leads to sum-spectral efficiencies beyond 30 bits per channel use.

\printbibliography

\end{document}